\documentclass[10pt]{article}
\usepackage[utf8]{inputenc}
\usepackage{amsmath,amsfonts,amssymb}
\usepackage[english]{babel}
\usepackage{hyperref}
\usepackage{geometry}
\usepackage{mathtools}

\renewcommand{\vec}[1]{\mathbf{#1}}
\newcommand{\x}{\vec{x}}

\title{Autonomous Experiments for Neutron Three-Axis Spectrometers (TAS) with Log-Gaussian Processes\thanks{Extended abstract for the virtual workshop \textit{Autonomous Discovery in Science and Engineering} organized by the Center for Advanced Mathematics for Energy Research Applications (CAMERA) from April 20--22, 2021. (\href{https://autonomous-discovery.lbl.gov}{https://autonomous-discovery.lbl.gov})}}

\author{M.~Teixeira Parente, G.~Brandl, C.~Franz, A.~Schneidewind, M.~Ganeva}

\date{\footnotesize
    Jülich Centre for Neutron Science at Heinz Maier-Leibnitz Zentrum (MLZ) \\
    Forschungszentrum Jülich GmbH \\
    Lichtenbergstraße 1, 85748 Garching, Germany \\
    \texttt{m.parente@fz-juelich.de}
}

\begin{document}
\maketitle

%%%%%%%%%%%%%%%%%%%%%%%%%%%%%%%%%%%%%%%%%%%%%%%
\section{Introduction}
%%%%%%%%%%%%%%%%%%%%%%%%%%%%%%%%%%%%%%%%%%%%%%%
Autonomous experiments are excellent tools to increase the efficiency of material discovery.
Indeed, AI and ML methods can help optimizing valuable experimental resources as, for example, beam time in neutron scattering experiments, in addition to scientists' knowledge and experience.

Active learning methods form a particular class of techniques that acquire knowledge on a specific quantity of interest by autonomous decisions on what or where to investigate next based on previous measurements.
For instance, Gaussian Process Regression (GPR) \cite{williams1996gaussian} is a well-known technique that can be exploited to accomplish active learning tasks for scattering experiments as was recently demonstrated \cite{noack2020advances,noack2020autonomous}.

\begin{figure}[h]
    \centering
    \includegraphics[width=\linewidth]{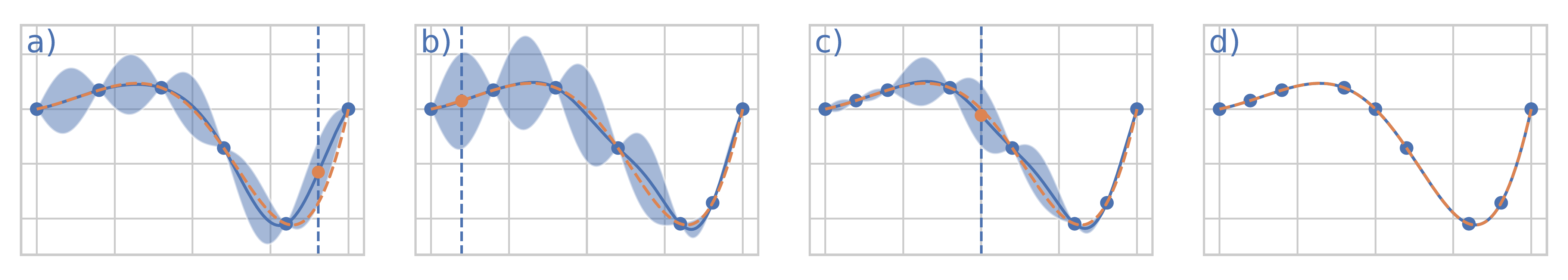}
    \caption{In images a)--d), the GPR posterior mean function (blue curve) iteratively approaches a function of interest (orange curve) by sampling at locations of highest uncertainty (orange dot in light blue area).}
    \label{fig:gpr_iter}
\end{figure}

Gaussian processes are not only capable to approximate functions by their posterior mean function, but can also quantify uncertainty about the approximation itself.
Hence, if we perform function evaluations at locations of highest uncertainty, the function can be ``optimally" learned in an iterative manner; see Fig.~\ref{fig:gpr_iter}.

We suggest the use of \textit{log-Gaussian processes}, being a natural approach to successfully conduct autonomous neutron scattering experiments in general and TAS experiments with the instrument PANDA at MLZ~\cite{schneidewind2015panda} in particular.

%%%%%%%%%%%%%%%%%%%%%%%%%%%%%%%%%%%%%%%%%%%%%%%
\section{Log-Gaussian processes}
%%%%%%%%%%%%%%%%%%%%%%%%%%%%%%%%%%%%%%%%%%%%%%%
From a mathematical perspective, GPR is done by conditioning a Gaussian process
\begin{equation}
    f \sim \mathcal{GP}(m(\x),k(\x,\x'))
\end{equation}
with mean function~$m(\x)$ and kernel function~$k(\x,\x')$ on noisy data
\begin{equation}
    \label{eq:data}
    d(\x_i) = f(\x_i) + e(\x_i)\eta
\end{equation}
at locations~$\x_i$, where $\eta\sim\mathcal{N}(0,1)$ is random noise and $e(\x_i)$ denotes its standard deviation at~$\x_i$.

In the following, we denote variables in $\mathbf{q}$-$\omega$ space by $\x=(\mathbf{q},\omega)^\top$.
There are two drawbacks in the application of standard GPR for scattering experiments.
First, since intensities are nonnegative, it is impossible that scattering functions are realizations of a Gaussian process which can take any real value.
However, it depends on the context whether that becomes a problem in particular situations.

Second, the main requirement in autonomous scattering experiments is to place the measurements in regions of signal and not in the background.
Thus, for the computation of ``optimal" measurement locations via maximizing an acquisition function, standard GPR has to rely on a reasonably good approximation for which a sufficiently large amount of data points is needed.

Utilizing log-Gaussian processes solves the first issue and at least weakens the second one.
We consider a log-Gaussian process
\begin{equation}
    \label{eq:lgp}
    g(\x) \coloneqq \exp(f(\x)),
\end{equation}
where $f\sim\mathcal{GP}(m(\x),k(\x,\x'))$ is a Gaussian process with mean function~$m(\x)$ and kernel function~$k(\x,\x')$.
This means that, in practice, we fit a Gaussian process~$f$ to the logarithm of the measured intensity data~$I(\x)$ and regard the corresponding log-Gaussian process; see Eq.~\eqref{eq:lgp}.
Its mean and variance become
\begin{align}
    \mathbf{E}[g(\x)] = \exp\left(m(\x)+\frac{\sigma^2(\x)}{2}\right)
    \text{ and }
    \mathbf{V}\text{ar}(g(\x)) = [\exp(\sigma^2(\x)-1)] \cdot \exp(2m(\x) + \sigma^2(\x)). \label{eq:var_lgp}
\end{align}
Then, the acquisition function to be maximized is simply
\begin{equation}
    \text{acq}(\x) = \mathbf{V}\text{ar}(g(\x)).
\end{equation}
Note that maximizing this function automatically places samples in regions of signal since the mean function~$m(\x)$ appears in~$\mathbf{V}\text{ar}(g(\x))$; see Eq.~\eqref{eq:var_lgp}.

\begin{figure}[ht]
    \centering
    \includegraphics[width=0.7\linewidth]{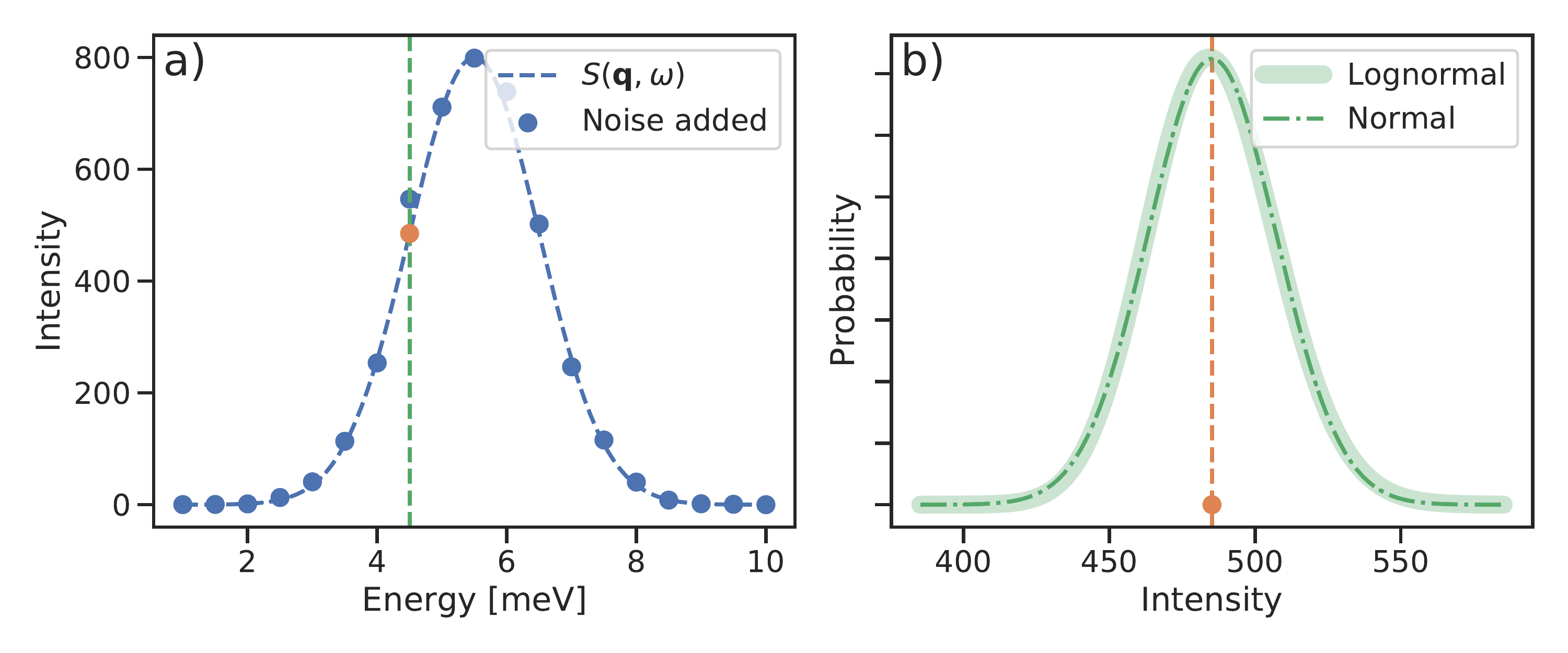}
    \caption{(a) Idealized constant-$\mathbf{q}$ scan data with Gaussian noise added to an assumed scattering function~$S(\mathbf{q},\omega)$.
        (b) Log-normal and normal noise distributions match for large intensities (here: $\approx480$).}
    \label{fig:noise_distr}
\end{figure}

However, this approach poses two immediate questions.
First, with log-Gaussian processes, noise on intensity data is assumed to be multiplicative log-Gaussian.
In contrast, from a physics perspective, the noise is assumed to be approximately additive Gaussian with standard deviation~$\sqrt{I(\x)}$ for sufficiently large intensities.
However, it turns out that a careful calculation of the noise variance~$e(\x)$ from Eq.~\eqref{eq:data} nearly gives equality of both noise distributions for large intensities; see Fig.~\ref{fig:noise_distr}.

The second question concerns the problem that this approach places more samples in regions with higher intensity than in other regions of interest with lower intensity.
We can circumvent this artefact by introducing a threshold for intensities considered and cut intensities higher than that threshold.
A reasonable way to determine such a threshold is subject of our current research.

\begin{figure}[ht]
    \centering
    \includegraphics[width=0.7\linewidth]{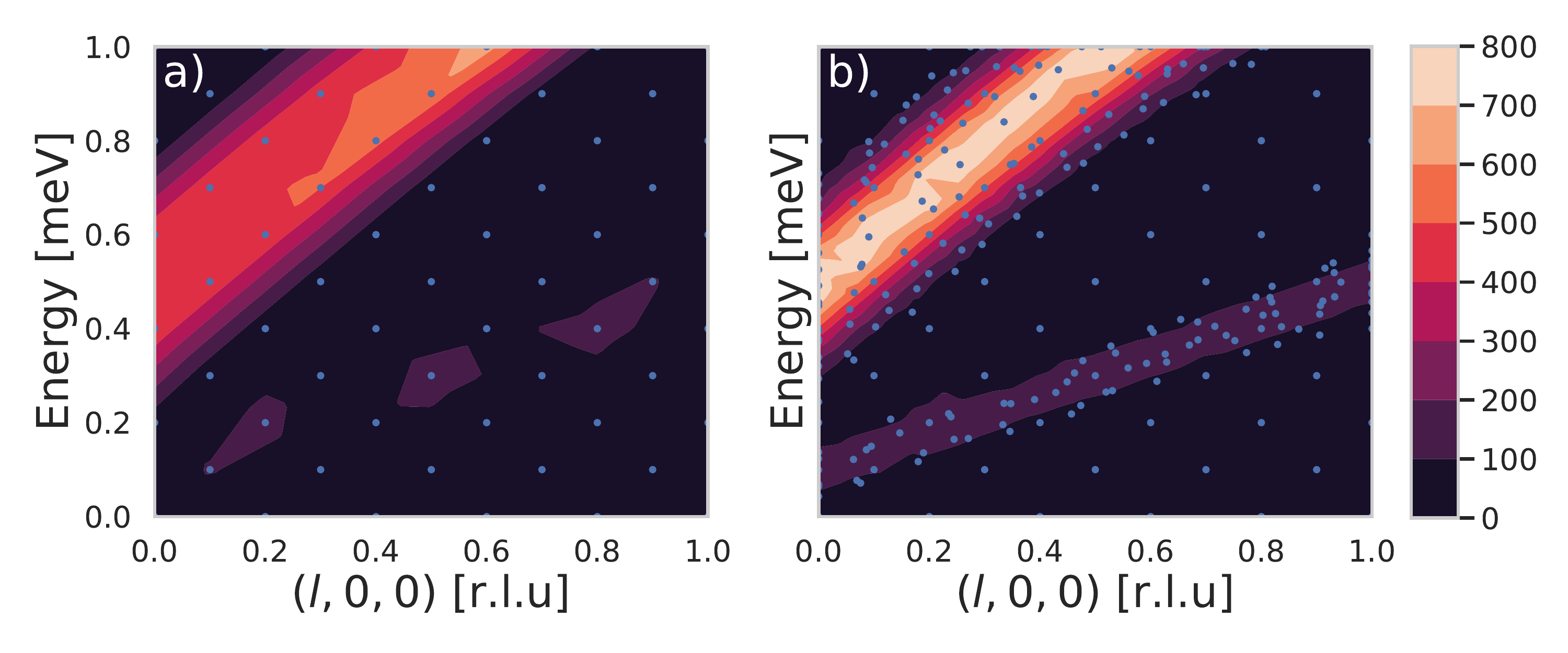}
    \caption{Blue dots represent measurement points and the color code provides the intensity of the signal.
        (a) Interpolation of initial measurements.
        (b) Interpolation of initial and suggested measurements.}
    \label{fig:separ_regio}
\end{figure}

Before applying neutrons, we have carried out several tests with synthetic data.
One of the test cases is shown in Fig.~\ref{fig:separ_regio}.
As can be seen in Fig.~\ref{fig:separ_regio}b, there are no unnecessary measurement points in the background after initialization and even the weaker intensity branch was found by the algorithm.

%%%%%%%%%%%%%%%%%%%%%%%%%%%%%%%%%%%%%%%%%%%%%%%
\section{Outlook}
%%%%%%%%%%%%%%%%%%%%%%%%%%%%%%%%%%%%%%%%%%%%%%%
The presented results serve as a starting point.
Our future research will focus on examining the robustness of our approach with respect to different scenarios and performance measures.
Indeed, to make the algorithms well suitable for wide scientific application, it is planned to collaborate with other interested groups, e.g., the CAMERA team (Marcus Noack), for setting up corresponding benchmarks to compare different approaches.
It is crucial to do this in close cooperation with the instrument scientists.

\section*{Acknowledgments}
This work is an outcome of the project \textit{Artificial Intelligence for Neutron and X-Ray Scattering} (AINX) funded by the  Helmholtz AI unit of the German Helmholtz Association which is greatly acknowledged by all authors.
We also thank Martin Boehm (ILL) and Marcus Noack (CAMERA) for fruitful discussions.

%%%%%%%%%%%%%%%%%%%%%%%%%%%%%%%%%%%%%%%%%%%%%%%
\bibliographystyle{abbrv}
\bibliography{refs}
%%%%%%%%%%%%%%%%%%%%%%%%%%%%%%%%%%%%%%%%%%%%%%%

\end{document}